\documentclass[aps,preprint,11pt,superscriptaddress,amsmath,amssymb]{revtex4}

\usepackage{graphicx} %
\usepackage{epstopdf} %
\usepackage{dcolumn} %
\usepackage{xcolor} %
\usepackage{amsmath,amssymb,amsfonts,mathrsfs,bm}%
\usepackage{hyperref} %
\usepackage[utf8]{inputenc} %
\usepackage[english]{babel} %
\usepackage[mathlines]{lineno} %
\usepackage{orcidlink} %

\usepackage[noabbrev,capitalise]{cleveref} %

\graphicspath{{figs/}} %

\newcommand{\re}{\ensuremath{\text{Re}}\xspace}

\newcommand{\DzToKspipi}{\ensuremath{\Dz\to\KS\pip\pim}\xspace}

\newcommand{\M}{\ensuremath{m(\KS\pip\pim)}\xspace}
\newcommand{\Q}{\ensuremath{Q}\xspace}

\newcommand{\pdf}{\ensuremath{P}}
\newcommand{\sig}{{\rm sig}}
\newcommand{\rnd}{{\rm rnd}}
\newcommand{\oth}{{\rm oth}}

\newcommand{\xFit}{4.0}
\newcommand{\xStat}{1.7}
\newcommand{\xSyst}{0.4}

\newcommand{\yFit}{2.9}
\newcommand{\yStat}{1.4}
\newcommand{\ySyst}{0.3}
\RequirePackage{xspace}

\newcommand{\CP}{\ensuremath{C\!P}\xspace}

\def\epem       {\ensuremath{e^+e^-}\xspace}

\def\qqbar {\ensuremath{q\overline q}\xspace}

\def\c     {\ensuremath{c}\xspace}

\def\pip   {\ensuremath{\pi^+}\xspace}
\def\pim   {\ensuremath{\pi^-}\xspace}

\def\Kbar  {\kern 0.2em\overline{\kern -0.2em K}{}\xspace}

\def\Kz    {\ensuremath{K^0}\xspace}
\def\Kzb   {\ensuremath{\Kbar^0}\xspace}
\def\KzKzb {\ensuremath{\Kz \kern -0.16em \Kzb}\xspace}
\def\Kp    {\ensuremath{K^+}\xspace}
\def\Km    {\ensuremath{K^-}\xspace}

\def\KpKm  {\ensuremath{\Kp \kern -0.16em \Km}\xspace}
\def\KS    {\ensuremath{K^0_{\scriptscriptstyle\rm S}}\xspace} 
\def\KL    {\ensuremath{K^0_{\scriptscriptstyle\rm L}}\xspace}

\def\Kstar   {\ensuremath{K^*}\xspace}

\def\D       {\ensuremath{D}\xspace}
\def\Dbar    {\kern 0.2em\overline{\kern -0.2em D}{}\xspace}

\def\Dz      {\ensuremath{D^0}\xspace}
\def\Dzb     {\ensuremath{\Dbar^0}\xspace}
\def\DzDzb   {\ensuremath{\Dz {\kern -0.16em \Dzb}}\xspace}
\def\Dp      {\ensuremath{D^+}\xspace}
\def\Dm      {\ensuremath{D^-}\xspace}

\def\DpDm    {\ensuremath{\Dp {\kern -0.16em \Dm}}\xspace}

\def\Dstarp  {\ensuremath{D^{*+}}\xspace}

\def\B       {\ensuremath{B}\xspace}
\def\Bbar    {\kern 0.18em\overline{\kern -0.18em B}{}\xspace}

\def\Bz      {\ensuremath{B^0}\xspace}
\def\Bzb     {\ensuremath{\Bbar^0}\xspace}
\def\BzBzb   {\ensuremath{\Bz {\kern -0.16em \Bzb}}\xspace}
\def\Bu      {\ensuremath{B^+}\xspace}
\def\Bub     {\ensuremath{B^-}\xspace}

\def\BpBm    {\ensuremath{\Bu {\kern -0.16em \Bub}}\xspace}

\def\Y#1S{\ensuremath{\Upsilon{(#1S)}}\xspace}%

\mathchardef\Deltares="7101
\mathchardef\Xi="7104
\mathchardef\Lambda="7103
\mathchardef\Sigma="7106
\mathchardef\Omega="710A

\def\Deltabar{\kern 0.25em\overline{\kern -0.25em \Deltares}{}\xspace}
\def\Lbar{\kern 0.2em\overline{\kern -0.2em\Lambda\kern 0.05em}\kern-0.05em{}\xspace}
\def\Sigbar{\kern 0.2em\overline{\kern -0.2em \Sigma}{}\xspace}
\def\Xibar{\kern 0.2em\overline{\kern -0.2em \Xi}{}\xspace}
\def\Obar{\kern 0.2em\overline{\kern -0.2em \Omega}{}\xspace}
\def\Nbar{\kern 0.2em\overline{\kern -0.2em N}{}\xspace}
\def\Xb{\kern 0.2em\overline{\kern -0.2em X}{}\xspace}

\newcommand{\tev}{\ensuremath{\mathrm{\,Te\kern -0.1em V}}\xspace}
\newcommand{\gev}{\ensuremath{\mathrm{\,Ge\kern -0.1em V}}\xspace}
\newcommand{\mev}{\ensuremath{\mathrm{\,Me\kern -0.1em V}}\xspace}
\newcommand{\kev}{\ensuremath{\mathrm{\,ke\kern -0.1em V}}\xspace}
\newcommand{\ev}{\ensuremath{\mathrm{\,e\kern -0.1em V}}\xspace}
\newcommand{\gevc}{\ensuremath{{\mathrm{\,Ge\kern -0.1em V\!/}c}}\xspace}
\newcommand{\mevc}{\ensuremath{{\mathrm{\,Me\kern -0.1em V\!/}c}}\xspace}
\newcommand{\gevcc}{\ensuremath{{\mathrm{\,Ge\kern -0.1em V\!/}c^2}}\xspace}
\newcommand{\mevcc}{\ensuremath{{\mathrm{\,Me\kern -0.1em V\!/}c^2}}\xspace}

\def\cm   {\ensuremath{{\rm \,cm}}\xspace}

\def\invfb   {\ensuremath{\mbox{\,fb}^{-1}}\xspace}

\def\mus  {\ensuremath{\rm \,\mus}\xspace}

\def\mus        {\ensuremath{\,\mu{\rm s}}\xspace}    %

\def\to                 {\ensuremath{\rightarrow}\xspace}

\def\gsim{{~\raise.15em\hbox{$>$}\kern-.85em
          \lower.35em\hbox{$\sim$}~}\xspace}
\def\lsim{{~\raise.15em\hbox{$<$}\kern-.85em
          \lower.35em\hbox{$\sim$}~}\xspace}

\usepackage{ifthen} %
\newboolean{articletitles}
\setboolean{articletitles}{true} %

\makeatletter%
\begin{document}
\includegraphics[width=3cm]{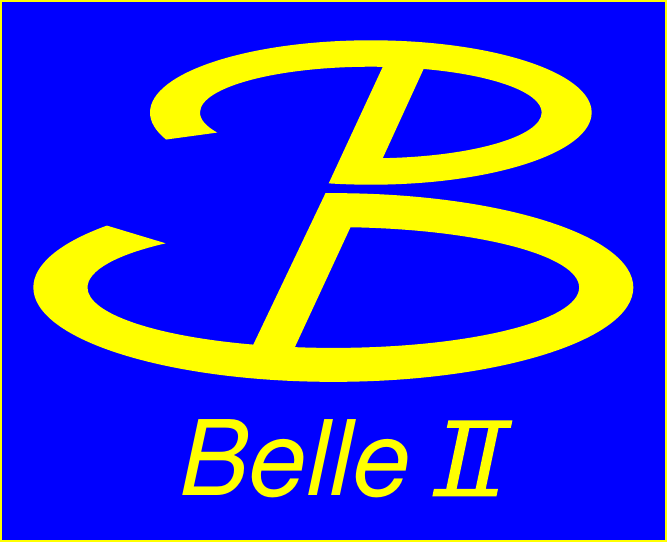}\vspace*{-1.9cm}

\begin{flushright}
Belle II Preprint 2024-027\\
KEK Preprint 2024-27
\end{flushright}\vspace{1.5cm}

\title{%
%
{\LARGE\bfseries\boldmath Model-independent measurement of \Dz-\Dzb mixing parameters in $\Dz\to\KS\pip\pim$ decays at Belle and Belle II}
 %
}
%
%
%
%
%
%
\collaboration{The Belle and Belle II Collaborations}
  \author{I.~Adachi\,\orcidlink{0000-0003-2287-0173},} %
  \author{L.~Aggarwal\,\orcidlink{0000-0002-0909-7537},} %
  \author{H.~Ahmed\,\orcidlink{0000-0003-3976-7498},} %
  \author{H.~Aihara\,\orcidlink{0000-0002-1907-5964},} %
  \author{N.~Akopov\,\orcidlink{0000-0002-4425-2096},} %
  \author{A.~Aloisio\,\orcidlink{0000-0002-3883-6693},} %
  \author{N.~Althubiti\,\orcidlink{0000-0003-1513-0409},} %
  \author{N.~Anh~Ky\,\orcidlink{0000-0003-0471-197X},} %
  \author{D.~M.~Asner\,\orcidlink{0000-0002-1586-5790},} %
  \author{H.~Atmacan\,\orcidlink{0000-0003-2435-501X},} %
% ...
%
%
 
\begin{abstract}
%
\noindent We perform a model-independent measurement of the \Dz-\Dzb mixing parameters using samples of \epem-collision data collected by the Belle and Belle~II experiments that have integrated luminosities of 951\invfb and 408\invfb, respectively. Approximately $2.05\times10^6$ neutral \D mesons are reconstructed in the $\Dz\to\KS\pip\pim$ channel, with the neutral \D flavor tagged by the charge of the pion in the $\Dstarp\to\Dz\pip$ decay. Assuming charge-parity symmetry, the mixing parameters are measured to be $ x = (\xFit\pm\xStat\pm\xSyst)\times 10^{-3} $ and $ y = (\yFit\pm\yStat\pm\ySyst)\times 10^{-3}$, where the first uncertainties are statistical and the second systematic. The results are consistent with previous determinations. 
 \end{abstract}

\maketitle
%
\section{Introduction}
Charm or \Dz-\Dzb mixing is the phenomenon in which neutral \D mesons oscillate into their anti-particles before decaying. It arises because the flavor eigenstates, \Dz and \Dzb, do not coincide with the mass eigenstates of the Hamiltonian. At quark level, mixing is induced by the exchange of $W^{\pm}$ bosons and down-type quarks or intermediate hadronic states, and its rate can be enhanced if particles beyond the standard model are also involved~\cite{summary:Lenz:2020awd}. Hence, precise measurements of charm mixing, and of charge-parity (\CP) violation in charm mixing, can serve as tools to probe new physics~\cite{newphysics:Isidori:2010kg}.

The mass eigenstates, $ D_1 $ and $ D_2 $, of neutral \D mesons can be expressed in terms of flavor eigenstates as
\begin{equation}
|D_{1,2}\rangle = p|\Dz\rangle  \pm  q |\Dzb\rangle\,,
\end{equation}
where $p$ and $q$ are complex parameters satisfying $|p|^2+|q|^2=1$. Following the convention from Ref.~\cite{pdg}, we define $\CP|\Dz\rangle = +|\Dzb\rangle$ such that, in the limit of \CP symmetry (i.e., $q=p$), $D_{1(2)}$ corresponds to the \CP-even (odd) eigenstate. The mixing parameters are defined as\footnote{This definition of the mixing parameters is consistent also with Ref.~\cite{hflav}. However, Ref.~\cite{hflav} uses different conventions for the definitions of $D_1$, $D_2$ and the \CP eigenvalues compared to Ref.~\cite{pdg}.}
\begin{equation}
x = \frac{m_1 - m_2}{\Gamma}\quad\text{and}\quad y =\frac{\Gamma_1 - \Gamma_2}{2\Gamma}\,,
\end{equation}
where $m_{1(2)}$ and $\Gamma_{1(2)} $ are the mass and the width of the $ D_{1(2)} $ state, and $\Gamma=(\Gamma_1+\Gamma_2)/2$ is the average decay-width. The average \Dz decay time is then given by $ \tau = 1/\Gamma$.

The world-average values of the mixing and \CP-violation parameters are $x = (4.07 \pm 0.44) \times 10^{-3} $, $y = (6.45^{+0.24}_{-0.23}) \times 10^{-3}$, $|q/p| = 0.994^{+0.016}_{-0.015}$, and $\arg(q/p) = (-2.6^{+1.1}_{-1.2})^\circ$~\cite{hflav}. These are the result of several experimental measurements performed over the past few decades and are currently dominated by results from LHCb~\cite{LHCb:2017uzt,LHCb:2021vmn,LHCb:2021ykz,LHCb:2022gnc}. The existence of charm mixing was first established in 2007 using a combination of results from BaBar and Belle~\cite{Aubert:2007wf,Staric:2007dt}. In 2012, LHCb observed charm mixing in a single experiment for the first time using $\Dz\to\Kp\pim$ decays\footnote{Inclusion of charge-conjugate processes is implied unless stated otherwise.}~\cite{LHCb:2012zll}, which significantly reduced the uncertainties in the world-average values of $x$ and $y$. This observation was later confirmed by CDF~\cite{CDF:2013gvz} and Belle~\cite{Belle:2014yoi} using the same decay mode. A nonzero value for $x$ has been reported by LHCb in 2021 using $\Dz\to\KS\pip\pim$ decays~\cite{LHCb:2021ykz}. Besides LHCb, Belle II is the only other experiment that has sufficient sensitivity to \Dz-\Dzb mixing to determine $x$ with precision below 1\%.

We measure the charm-mixing parameters $x$ and $y$ using $\Dz\to\KS\pip\pim$ decays reconstructed in the Belle and Belle II datasets, which have integrated luminosities of 951\invfb and 408\invfb, respectively. The analysis uses $\Dz\to\KS\pip\pim$ decays originating from $\Dstarp\to\Dz\pip$ decays so that the production flavor of the neutral \D mesons can be determined from the charge of the accompanying pions. Signal decays are separated from background using fits to the two-dimensional distribution of $\KS\pip\pim$ mass, \M, and energy released in the \Dstarp decay, \Q. The decay-time distribution of the signal candidates is described using a method that does not rely on precise modeling of the $\Dz\to\KS\pip\pim$ decay amplitude, which eliminates model-dependent systematic uncertainties~\cite{Bondar:2010qs,Thomas:2012qf}. This model-independent method builds on the same ideas developed to measure the CKM angle $\gamma$ from $B^-\to\D(\to \KS\pip\pim)K^-$ decays, known as the BPGGSZ method~\cite{Giri:2003ty,Bondar:2005ki,Bondar:2008hh}. By partitioning the Dalitz plot into bins, the need for an explicit amplitude model is avoided, and the decay-time distribution depends on a small number of hadronic parameters that encode the relevant dynamics, in addition to the mixing parameters. The hadronic parameters are measured with sufficient precision at charm factories where pairs of $\Dz\!\Dzb$ mesons are coherently produced in \epem collisions at the $\psi(3770)$ resonance~\cite{Libby:2010nu,strongphase:BESIII:2020hlg,strongphase:BESIII:2020khq}. Therefore, a simultaneous fit to the decay-time distributions of all Dalitz-plot bins, in which the external information on the hadronic parameters is used as a constraint, gives access to the mixing parameters. To avoid experimenter's bias while developing the analysis, the mixing parameters in the fit were shifted by unknown offsets randomly sampled between $-2\times10^{-2}$ and $+2\times10^{-2}$. The offsets were revealed only after having finalized the analysis procedure and evaluation of uncertainties. 

The paper is structured as follows. The formalism of the model-independent method is discussed in \cref{sec:method}. The Belle and Belle II detectors are presented in \cref{sec:detector}. The data and simulation samples used in the analysis are described in \cref{sec:datasets}. \Cref{sec:selection} reports the reconstruction and selection of the signal decays, and the resulting sample composition. The time-dependent mixing fit is presented in \cref{sec:time-dependent-fit}. Systematic uncertainties are evaluated in \cref{sec:systematics}. A summary of the results is given before concluding.

\section{Formalism\label{sec:method}}
We use the Dalitz-plot  formalism~\cite{Dalitz:1953cp,Fabri:1954zz} to parameterize the $\Dz\to\KS\pip\pim$ decay amplitude. To simplify the simultaneous treatment of \Dz and \Dzb decays, we build the Dalitz plot using the following flavor-dependent definition of squared invariant masses:
\begin{equation}
m_\pm^2 \equiv \left\{\begin{aligned}
m^2(\KS\pi^\pm)\quad & \text{for initially produced \Dz mesons}\\
m^2(\KS\pi^\mp)\quad & \text{for initially produced \Dzb mesons}\\
\end{aligned}\right.\,.
\end{equation}
The decay-time rate of initially produced \Dz and \Dzb mesons decaying to the $\KS\pip\pim$ final state are
\begin{align}
\left|{T}(m_+^2,m_-^2;t)\right|^2 &= \left|A(m_+^2,m_-^2)\, g_+(t) + \bar{A}(m_-^2,m_+^2)\, \frac{q}{p}\, g_-(t)\right|^2\,, \label{eq:rate-D0}\\
\left|\overline{T}(m_+^2,m_-^2;t)\right|^2 &= \left|\bar{A}(m_+^2,m_-^2)\, g_+(t) + A(m_-^2,m_+^2)\, \frac{p}{q}\, g_-(t)\right|^2\,,\label{eq:rate-D0bar}
\end{align}
where $A(m_+^2,m_-^2)$ and $\bar{A}(m_+^2,m_-^2)$ indicate the \Dz and \Dzb decay amplitudes as a function of the Dalitz-plot coordinates, and
\begin{equation}
g_\pm(t) = \theta(t)e^{-im  t}e^{-\Gamma t/2}~^{\cosh}_{\sinh}(z\Gamma t/2)\,,
\end{equation}
where $m=(m_1+m_2)/2$ is the average mass of neutral \D mesons, $\theta$ is the Heaviside function, and $z$ equals $-(y+ix)$.

\begin{figure}[t]
	\centering
	\includegraphics[width=0.5\textwidth]{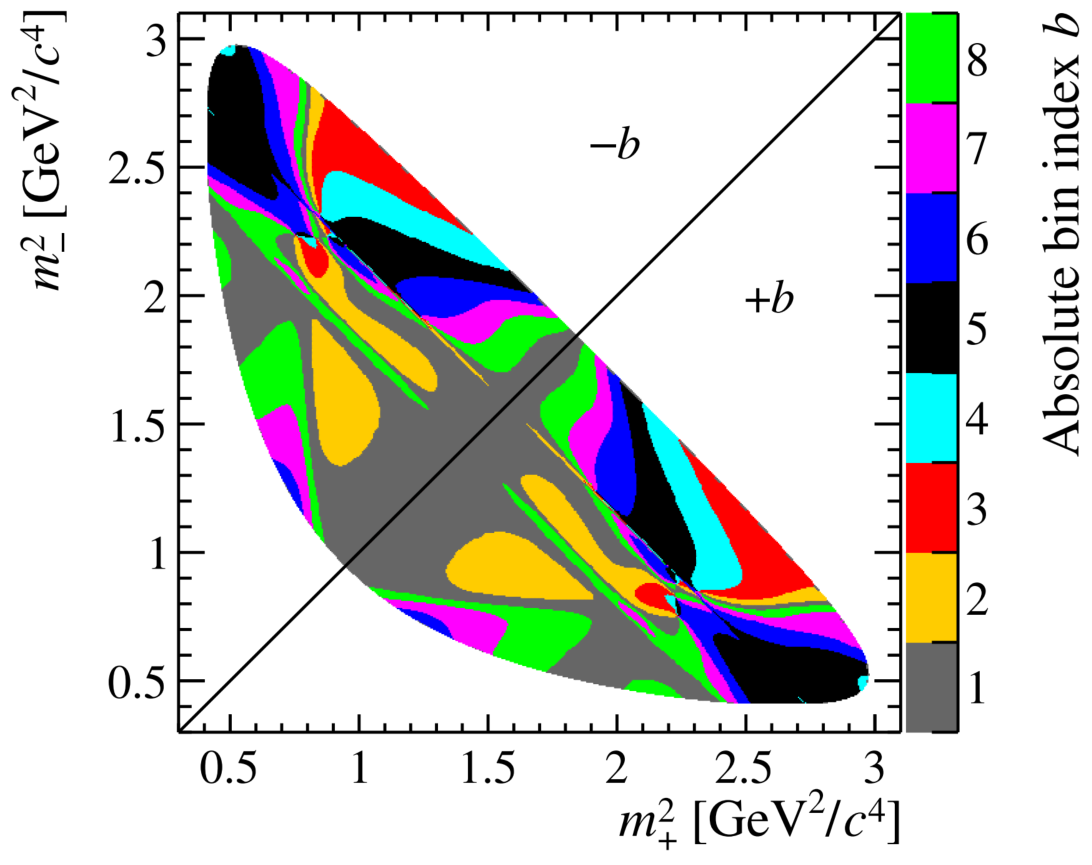}
	\caption{Iso-$\Delta\delta$ binning of the $\Dz\to\KS\pip\pim$ Dalitz plot, developed by CLEO~\cite{Libby:2010nu} using the amplitude model from Ref.~\cite{BaBar:2008inr}. The bins are symmetric with respect to the $m_+^2=m_-^2$ bisector; positive indices refer to bins in the (lower) $m_+^2 > m_-^2$ region; negative indices refer to those in the (upper) $m_+^2 < m_-^2$ region. Colors indicate the absolute value of the bin index $b$.\label{fig:equalbinning}}
\end{figure}

A model-dependent measurement of the mixing parameters relies on fitting \cref{eq:rate-D0,eq:rate-D0bar} to the data with a model for the variation of the decay amplitudes across the Dalitz plot. The amplitude model introduces large and difficult-to-estimate systematic uncertainties, which can limit the precision of the measurement~\cite{CLEO:2005qym,CLEO:2002uvu,Belle:2007tti,delAmoSanchez:2010xz,Peng:2014oda}. In this work, we avoid the dependence on the amplitude model by splitting the data into $n$ pairs of Dalitz-plot bins symmetric with respect to the diagonal $m_+^2=m_-^2$. In this analysis, we use the ``iso-$\Delta\delta$'' $n=8$ binning scheme of the Dalitz plot~\cite{Libby:2010nu} shown in \cref{fig:equalbinning}. We then define
\begin{align}
F_b & = \int_b\left|A(m^2_+,m^2_-)\right|^2dm^2_+dm^2_-\,, \\ 
\bar{F}_b & = \int_b\left|\bar{A}(m^2_+,m^2_-)\right|^2dm^2_+dm^2_-\,, \\
X_b & =  \frac{1}{\sqrt{F_b\bar{F}_{-b}}}\int_bA^*(m^2_+,m^2_-)\bar{A}(m^2_-,m^2_+)dm^2_+dm^2_-\,,
\end{align}
where $b=-n,...,-1,+1,...,+n$ is the index of the Dalitz-plot bin. Positive and negative indices indicate bins in the Dalitz-plot semispaces $m_+^2>m_-^2$ and $m_-^2>m_+^2$, respectively. Here, $F_b$ and $\bar{F}_b$ are event yields in the Dalitz-plot bin $b$ at $t=0$ and can be determined directly when fitting the data. The hadronic parameters $X_b=c_b-is_b$ quantify the interference between the \Dz and \Dzb amplitudes. The imaginary and real parts of $X_b$ are measured in experiments where $\Dz\!\Dzb$ pairs are produced in an entangled state~\cite{strongphase:BESIII:2020hlg,strongphase:BESIII:2020khq} and constrained in the fit to the data, thus eliminating the need for an amplitude model.

Assuming \CP symmetry in both decay and mixing means $A(m^2_+,m^2_-) = \bar{A}(m^2_+,m^2_-)$ and $q=p$. The phase difference between $A(m^2_+,m^2_-) $ and $ \bar{A}(m^2_-,m^2_+) $, $\Delta\delta(m^2_+,m^2_-)$, is then only due to the strong interaction. In this case, $ F_b = \bar{F}_b $, $ X_b = \bar{X}_b $ and $ X_b = X^*_{-b}$. Hence, integrating \cref{eq:rate-D0,eq:rate-D0bar} in Dalitz-plot bin $ b $ or $ -b $ results in the following relations 
\begin{align}
p_b(t) &= \int_b\left|{T}_f(m_+^2,m_-^2;t)\right|^2dm_+^2dm_-^2 = \int_b\left|\overline{T}_{\!f}(m_+^2,m_-^2;t)\right|^2dm_+^2dm_-^2\nonumber \\
& \propto g^2_+(t) + r_{b}g^2_-(t) + 2\sqrt{r_b}\,\re[X_bg^*_+(t)g_-(t)] \ ,\label{eq:rate-D0-pb} \\
p_{-b}(t) &= \int_{-b}\left|{T}_f(m_+^2,m_-^2;t)\right|^2dm_+^2dm_-^2 = \int_{-b}\left|\overline{T}_{\!f}(m_+^2,m_-^2;t)\right|^2dm_+^2dm_-^2 \nonumber\\
& \propto r_{b}g^2_+(t) + g^2_-(t) + 2\sqrt{r_b}\,\re[X_{-b}g^*_+(t)g_-(t)] \,,\label{eq:rate-D0-mb}
\end{align}
where $ r_b=F_{-b}/F_{b}$.

The probability density functions (PDFs) used in the fit to the data are based on \Cref{eq:rate-D0-pb,eq:rate-D0-mb}, after having included experimental effects such as detector resolution and contributions from background processes which are discussed in \cref{sec:time-dependent-fit}.

\section{Belle and Belle II detectors\label{sec:detector}}
The Belle experiment~\cite{Belle:2000cnh,Belle:2012iwr} operated at KEKB asymmetric-energy $\epem$ collider~\cite{KEKB:Abe:2013kxa,KEKB:Kurokawa:2001nw} between 1999 and 2010. The detector consisted of a large-solid-angle spectrometer, which included a double-sided silicon-strip vertex detector, a 50-layer central drift chamber, an array of aerogel threshold Cherenkov counters, a barrel-like arrangement of time-of-flight scintillation counters, and an electromagnetic calorimeter comprised of CsI(Tl) crystals. All subdetectors were located inside a superconducting solenoid coil that provided a 1.5\,T magnetic field. An iron flux-return yoke, placed outside the coil, was instrumented with resistive-plate chambers to detect \KL mesons and identify muons.  Two inner detector configurations were used: a 2.0\cm radius beam pipe and a three-layer silicon vertex detector (SVD1); and, from October 2003, a 1.5\cm radius beam pipe, a four-layer silicon vertex detector, and a small-inner-cell drift chamber (SVD2)~\cite{Natkaniec:2006rv}.

The Belle II detector~\cite{Abe:2010gxa} is an upgrade with several new subdetectors designed to handle the significantly larger beam-related backgrounds of the new collider, SuperKEKB~\cite{Akai:2018mbz}. It consists of a silicon vertex detector wrapped around a 1\cm radius beam pipe and comprising two inner layers of pixel detectors and four outer layers of double-sided strip detectors, a 56-layer central drift chamber, a time-of-propagation detector, an aerogel ring-imaging Cherenkov detector, and an electromagnetic calorimeter, all located inside the same solenoid as used for Belle. The flux return outside the solenoid is instrumented with resistive-plate chambers, plastic scintillator modules, and an upgraded read-out system to detect muons, \KL mesons, and neutrons. For the data used in this paper, collected between 2019 and 2022, only part of the second layer of the pixel detector, covering 15\% of the azimuthal angle, was installed.

\section{Data sets\label{sec:datasets}}
This analysis uses $ \Dstarp\to\Dz(\to\KS\pip\pim)\pip $ candidates reconstructed in Belle and Belle~II data. The Belle data samples were collected on or near the \Y4S and \Y5S resonances and have integrated luminosities of 800\invfb (with 80\% of this taken in the SVD2 configuration) and 151\invfb, respectively~\cite{Belle:2012iwr}. The Belle II data sample, collected near the \Y4S resonance, has an integrated luminosity of 408\invfb~\cite{lumi}.

We use simulation to identify sources of background, quantify reconstruction effects, determine fit models, and validate the analysis procedure. We generate $\epem\to\Upsilon(nS)$ ($n=4,5$) events and simulate particle decays with \textsc{EvtGen}~\cite{Lange:2001uf}; we generate continuum $\epem\to\qqbar$ (where $q$ is a $u$, $d$, $c$, or $s$ quark) with \textsc{Pythia6}~\cite{Sjostrand:2006za} for Belle, and with \textsc{KKMC}~\cite{Jadach:1990mz} and \textsc{Pythia8}~\cite{Sjostrand:2014zea} for Belle~II; we simulate final-state radiation with \textsc{Photos}~\cite{Barberio:1990ms,Barberio:1993qi}; we simulate detector response using \textsc{Geant3}~\cite{Brun:1073159} for Belle and \textsc{Geant4}~\cite{Agostinelli:2002hh} for Belle II. In the Belle simulation, beam backgrounds are taken into account by overlaying random-trigger data. In the Belle II simulation, they are accounted for by simulating the Touschek effect~\cite{PhysRevLett.10.407}, beam-gas scattering, and luminosity-dependent backgrounds from Bhabha scattering and two-photon quantum-electrodynamic processes~\cite{Lewis:2018ayu,Natochii:2023thp}.

\section{Event selection and sample composition\label{sec:selection}\label{sec:mq-fit}}
We use the Belle II analysis software framework (basf2) to reconstruct both Belle and Belle II data~\cite{Kuhr:2018lps,basf2-zenodo}. The Belle data are converted to the Belle II format for basf2 compatibility using the B2BII framework~\cite{Gelb:2018agf}. 

Events are selected by a trigger according to either the total energy deposited in the electromagnetic calorimeter or the number of charged-particle tracks reconstructed in the central drift chamber. The efficiency of the trigger selection is found to be close to 100\% for events containing signal $\Dstarp\to\Dz(\to\KS\pip\pim)\pip$ candidate decays.

Signal candidates are reconstructed starting from combinations of two oppositely charged pions to form $ \KS\to\pip\pim$ candidates. A fit to determine the \KS decay-vertex position is performed and the resulting $\pip\pim$ mass is required to be in the range $[0.488,0.508]\gevcc$. The \KS candidates are then combined with two oppositely charged pions, having small radial and longitudinal distances of closest approach to the interaction region ($\Delta r <1 \cm$ and $|\Delta z|<5 \cm$), to form \DzToKspipi candidates. The \Dz candidate is then combined with another charged pion satisfying the same track-quality requirements to form $\Dstarp\to\Dz\pip$ candidates. Due to the small energy released in the decay, the pion from the \Dstarp decay has much lower momentum than the pions from the \Dz decay and is therefore identified as the ``soft'' pion in the following. To suppress the candidates arising from the decay of a \B meson, the momentum of \Dstarp in the \epem center-of-mass frame is required to be larger than $2.5\gevc$ ($ 3.1\gevc $) for the data collected near the \Y4S (\Y5S) resonance. A kinematic-vertex fit~\cite{treefitter} to the selected \Dstarp candidates determines the \Dz and \Dstarp decay vertices while constraining the reconstructed $\pip\pim$ mass to the known \KS mass~\cite{pdg} and the \Dstarp decay-vertex to the measured position of the \epem interaction region. Candidates with failed fits or $\chi^2/\text{ndf}>200/10$ are rejected. This fit calculates the \KS flight length $L$, i.e. the distance between the \KS and \Dz decay vertices, and its uncertainty $\sigma_L$. The flight-length significance $L/\sigma_L$ is required to be larger than $20$ to suppress background candidates having no \KS meson in the \Dz final state, such as $\Dz\to\pip\pim\pip\pim$ decays. The \Dstarp candidates are required to satisfy $1.8<\M<2.0 \gevcc$ and $0.2<\Q<20.0 \mev$.

To ensure that all candidates lie in the kinematically allowed phase-space region, the vertex fit is re-run with an additional \Dz-mass constraint. The fit never fails and no candidates are removed in this step. We use the results of this second fit to compute the Dalitz-plot coordinates. All other quantities, including the \Dz decay time $t$ and its uncertainty $\sigma_t$, are computed with the result of the fit without the \Dz-mass constraint.

In 15\% of events more than one \Dstarp candidate is selected. If the multiple candidates result from the
combination of final-state particles reconstructed from cloned tracks, we remove them. Otherwise, we accept all candidates. This procedure removes approximately 1\% of selected candidates and has been verified to not introduce a bias.

Selected candidates are categorized as signal, random-pion background, and other background. Signal candidates are those for which the full $\Dstarp\to\Dz(\to\KS\pip\pim)\pip$ decay chain is correctly reconstructed. The signal candidates feature narrow peaks in both \M and \Q distributions. Random-pion background candidates are defined as those in which the \Dz is correctly reconstructed but associated to an unrelated soft pion to form the \Dstarp candidate. They peak exactly as the signal in \Dz mass, but are smoothly distributed in \Q value. The remaining candidates are referred to as other background candidates. They have smooth \M and \Q distributions.

The fractions of the three components in each Dalitz-plot bin are determined from an unbinned maximum-likelihood fit to the two-dimensional distribution of \M versus \Q. The signal \M distribution is modeled with a Crystal Ball function~\cite{Gaiser:Phd,Skwarnicki:1986xj} and two Gaussian functions, with shared mean value. The \Q distribution of signal is modeled with a Johnson unbounded distribution~\cite{johnson} and two Gaussian functions. The width of the second Gaussian function has a quadratic dependence on \M. For the random-pion component, the \M distribution is identical to the signal component, while the \Q distribution is modeled with the two-body phase-space function $Q^{1/2} + \alpha Q^{3/2} + \beta Q^{5/2}$. The distributions of the other background are modeled with a third-order polynomial for \M and a two-body phase-space function for \Q.

\begin{figure*}[t]
\centering
\includegraphics[width=\textwidth]{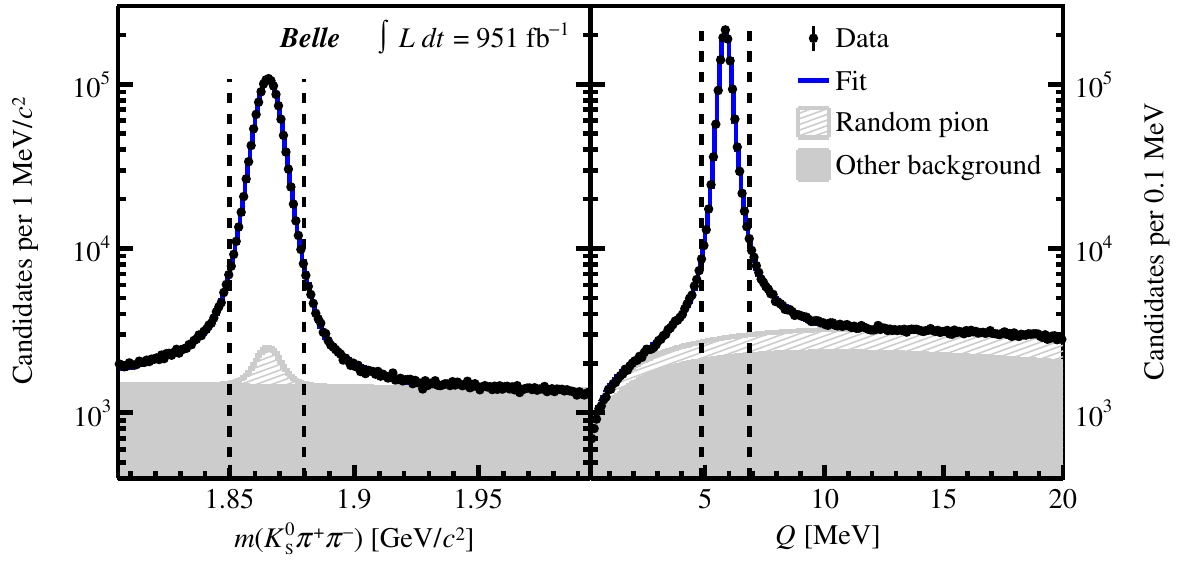}\\
\includegraphics[width=\textwidth]{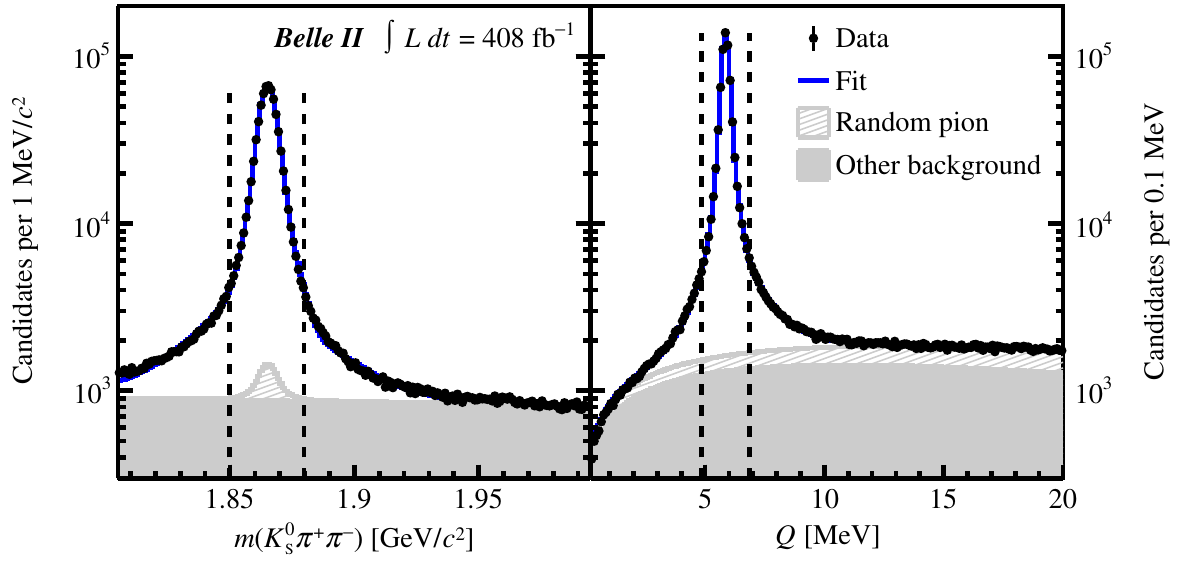}\\
\caption{Distributions of (left) \M for candidates populating the \Q signal region and (right) of \Q for candidates populating the \M signal region, in (top) Belle and (bottom) Belle II data with fit projections overlaid. The signal regions are indicated with vertical lines.  \label{fig:mqfit}}
\end{figure*}

\Cref{fig:mqfit} shows the distributions of \M and \Q for the candidates in the respective signal regions, with fit projections overlaid. The \M signal region is defined by $ |\M - m_{\Dz}| < 15 \mevcc$, where $ m_{\Dz} $ is the known \Dz mass~\cite{pdg}. The \Q signal region corresponds to the range $[4.85,6.85]\mev$. These distributions are integrated over all Dalitz-plot bins. We reconstruct approximately $1.35\times10^6$ and $0.70\times10^6$ signal candidates in Belle and Belle II, respectively. The signal purity in the signal region is approximately 96\% for both samples.

\begin{figure*}[t]
\centering
\includegraphics[width=0.5\textwidth]{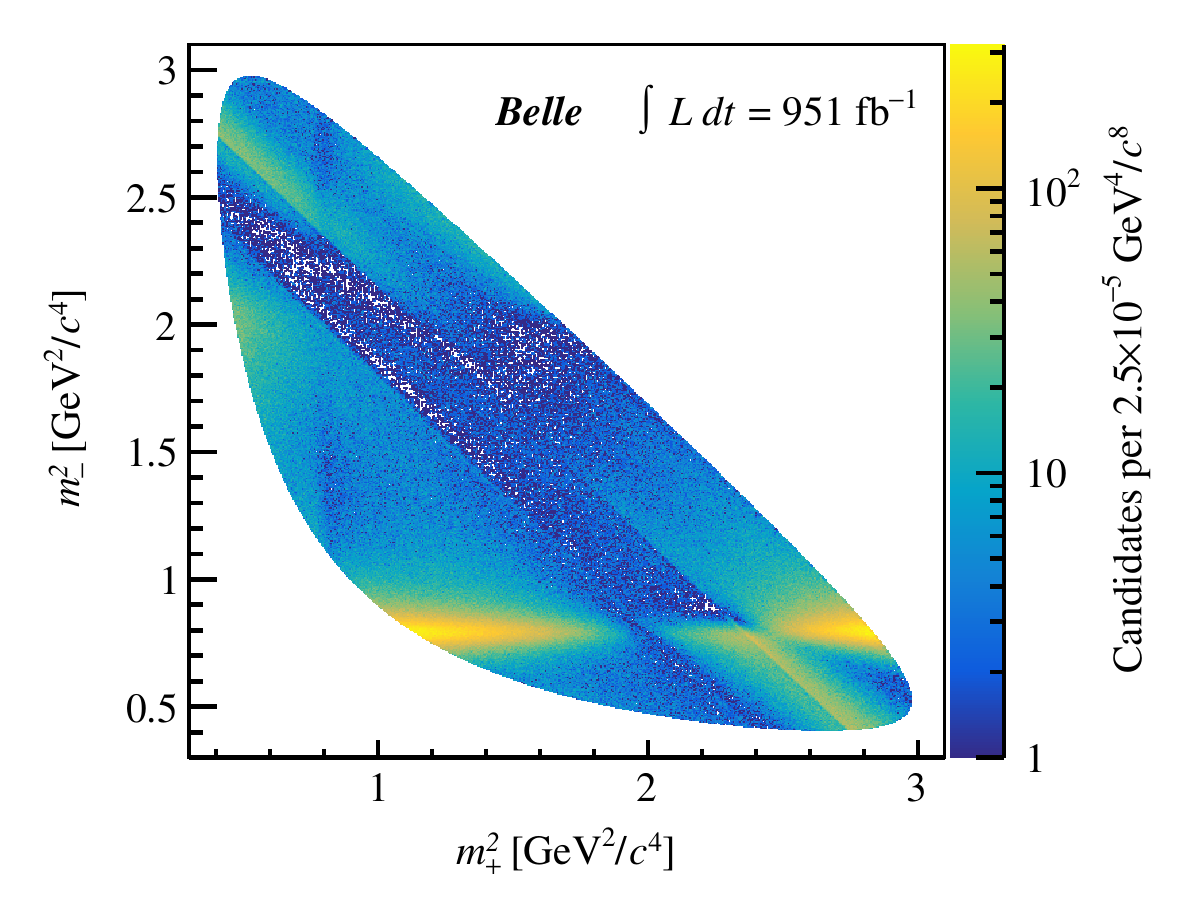}\hfil
\includegraphics[width=0.5\textwidth]{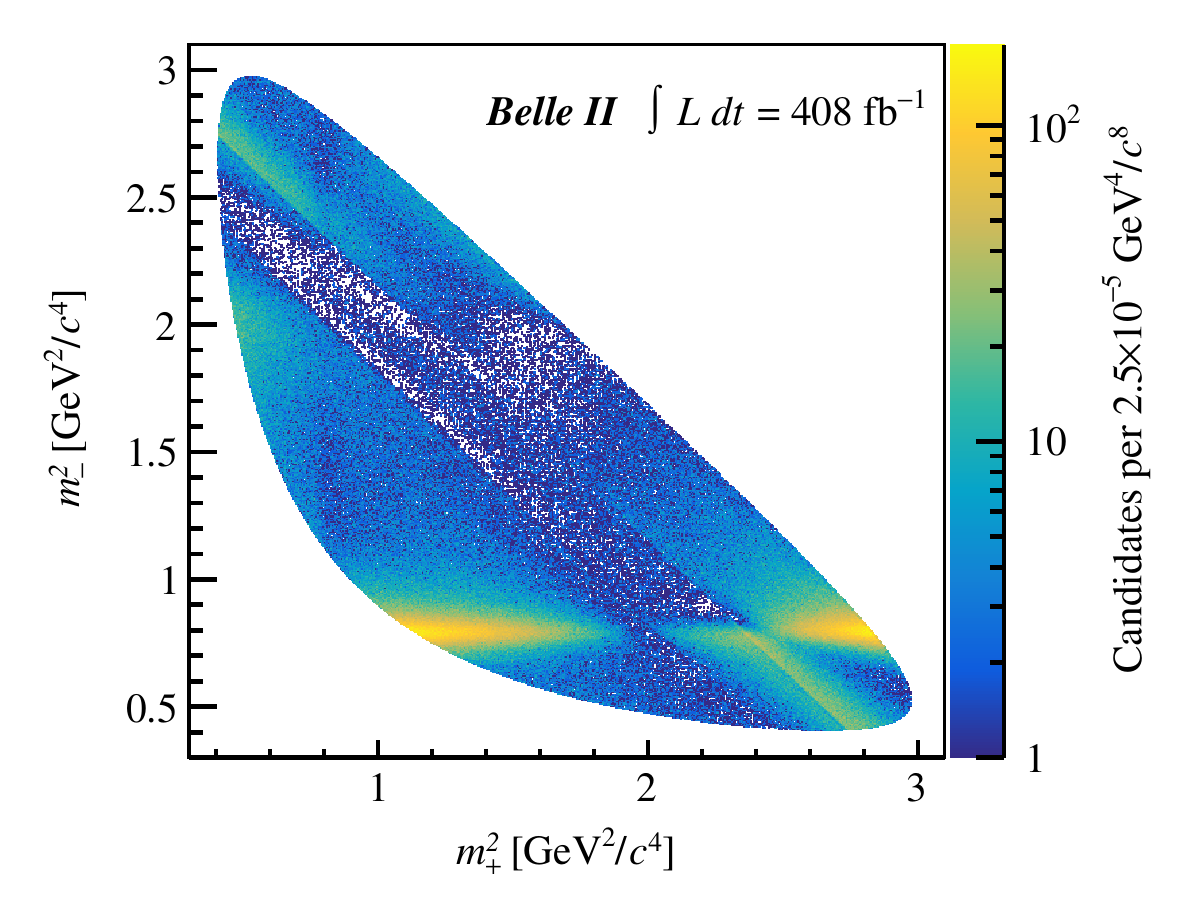}\\
\caption{Dalitz plots of candidates populating the signal region in (left) Belle and (right) Belle II data. \label{fig:dalitz_plot_data.main}}
\end{figure*}

The Dalitz plots of the candidates populating the combined \M and \Q signal region is shown in \cref{fig:dalitz_plot_data.main}. The structures due to the dominant intermediate processes are clearly visible~\cite{delAmoSanchez:2010xz,Peng:2014oda}. These include the Cabibbo-favored $\Dz\to\Kstar(892)^-\pip$ amplitude populating the horizontal band in the $m_+^2>m_-^2$ region (corresponding to the positive Dalitz-plot bins), the \CP-odd $\Dz\to\KS\rho(770)^0$ amplitude appearing as a band orthogonal to the $m_+^2=m_-^2$ diagonal, and the $\Dz\to\KS(\pip\pim)_{S\text{-wave}}$ component, which also includes  the $f_0(980)$ resonance. The vertical band in the region $m_+^2<m_-^2$ (corresponding to the negative Dalitz-plot bins) include doubly Cabibbo-suppressed $\Dz\to\Kstar(892)^+\pim$ decays, mixed \D mesons followed by the Cabibbo-favored $\Dzb\to\Kstar(892)^+\pim$ decay, and Cabibbo-favored $\Dzb\to\Kstar(892)^+\pim$ decays mistagged as \Dz decays by a random soft pion.

\section{Time-dependent mixing fit}\label{sec:time-dependent-fit}
The mixing parameters are determined using an unbinned maximum-likelihood fit to the $(t, \sigma_t) $ distributions of the candidates populating the signal region and split into the 16 Dalitz-plot bins and into four different data subsets. The subsets correspond to Belle data collected with SVD1, Belle \Y4S data collected with SVD2, Belle \Y5S data, and Belle II data.

The PDF of the signal decays is constructed from \cref{eq:rate-D0-pb,eq:rate-D0-mb} by including reconstruction effects. However, we neglect effects due to nonuniform efficiency variations and mass resolutions across the Dalitz plot, which are accounted for in the systematic uncertainties (\cref{sec:systematics}), and only include the effect of the decay-time resolution. We model the decay-time resolution using the per-candidate decay-time uncertainty $ \sigma_t $. The two-dimensional $(t,\sigma_t)$ PDF of the signal candidates is expressed as the product between the PDF of $\sigma_t$ and the PDF of $t$ given the value of $\sigma_t$. The latter is expressed as the convolution of the decay rate of \cref{eq:rate-D0-pb,eq:rate-D0-mb} with the resolution function $ R(t|\sigma_t) $, which depends on $ \sigma_t$:
\begin{equation}\label{eqn:pdf-sig}
\pdf_{\sig}(t,\sigma_t|b)=\pdf_{\sig}(t|\sigma_t,b)\pdf_{\sig}(\sigma_t|b) \propto \left[p_{b}(t) \otimes R(t|\sigma_t)\right]\pdf_{\sig}(\sigma_t|b) \,,
\end{equation}
where $\pdf_{\sig}(\sigma_t|b)$ is a histogram template determined, independently for each bin $b$ and for each data subset, by subtracting the $\sigma_t$ distribution of the candidates in the \M sideband $[1.97,2.00]\gevcc$ from the $\sigma_t$ distribution of the signal region using the measured background fraction. The resolution function $ R(t|\sigma_t) $ is parameterized as a double Gaussian distribution
\begin{equation}\label{eqn:resolution}
R(t|\sigma_t) = f G(t| \mu, \sigma_1) + (1-f) G(t|\mu,s_2 \sigma_1)\,,
\end{equation}
where $\sigma_1 = s_1\sigma_t + s_{11}\sigma_t^2$. The parameters $f$, $\mu$,  $s_1$, $s_{11}$, and $s_2$ are determined directly from the fit to the data (in the fit to Belle data $s_{11}$ is fixed to zero). Independent parameters are considered for each data subset. We then obtain
\begin{equation}\label{eqn:pdf_sig_t_sigma_t}
\pdf_{\sig}(t|\sigma_t,b) = \frac{k_+(t) + r_{b}k_-(t) + 2\sqrt{r_{b}}\,\re[X_bk_{+-}(t)]}
{K_+ + r_{b}K_- + 2\sqrt{r_{b}}\,\re(X_bK_{+-})} \,.
\end{equation}
The functions $ k_+(t)$, $ k_-(t)$ and $ k_{+-}(t)$ are, respectively, the convolutions of $ g^2_{+}(t)$, $ g^2_{-}(t)$ and $ g^*_{+}(t)g_{-}(t) $ and the resolution function $R(t|\sigma_t)$. They are expressed analytically using the following relations~\cite{Karbach:2014qba}:
\begin{align}
g^2_+(t) \otimes G(t| \mu, \sigma)
&= \frac{1}{4}\bigl[\psi(\chi,\kappa_+) + \psi(\chi,\kappa_{+i}) + \psi(\chi,\kappa_{-i}) + \psi(\chi,\kappa_-)\bigr] \,, \\
g^2_-(t) \otimes G(t| \mu, \sigma)
&= \frac{1}{4}\bigl[\psi(\chi,\kappa_+) - \psi(\chi,\kappa_{+i}) - \psi(\chi,\kappa_{-i}) + \psi(\chi,\kappa_-)\bigr]\,, \\
g_+^*(t)g_-(t) \otimes G(t| \mu, \sigma)
&= \frac{1}{4}\bigl[\psi(\chi,\kappa_+) + \psi(\chi,\kappa_{+i}) - \psi(\chi,\kappa_{-i}) - \psi(\chi,\kappa_-)\bigr]\,,
\end{align}
with
\begin{gather}
\chi = \frac{t-\mu}{\sigma},\quad \kappa_{\pm}=(1\pm y)\Gamma \sigma,\quad \kappa_{\pm i}=(1\pm ix)\Gamma \sigma, \nonumber\\
\psi(\chi, \kappa) = \frac{1}{2}e^{\frac{\kappa^2-2\chi \kappa}{2}}\left[1+{\rm erf}\left(\frac{\chi-\kappa}{\sqrt{2}}\right)\right] \,.
\end{gather}
The integrals of $ k_+(t)$, $ k_-(t)$ and $ k_{+-}(t)$ over decay time are
\begin{equation}
K_+ = \int_{-\infty}^{\infty} k_+(t)dt, \quad K_- = \int_{-\infty}^{\infty} k_-(t)dt, \quad K_{+-} = \int_{-\infty}^{\infty}k_{+-}(t)dt\,. 
\end{equation}

The total PDF includes the contribution of the backgrounds. For a given bin $b$, it is defined as
\begin{equation}\label{eqn:mixing-fit-signal-and-background}
\pdf(t,\sigma_t|b) = \mathcal{C}_b \left[ f_{\sig}^b\pdf_{\sig}(t,\sigma_t|b)
+ f_{\rnd}^b\pdf_{\rnd}(t,\sigma_t|b) + (1-f_{\sig}^b-f_{\rnd}^b)\pdf_{\oth}(t,\sigma_t|b)\right]\,,
\end{equation}
with
\begin{equation}
\mathcal{C}_b = \left\{\begin{aligned}
\frac{1}{1+C_b} \quad b>0\\
\frac{C_{b}}{1+C_{b}} \quad b<0\\
\end{aligned}\right.\,.  
\end{equation}
Here, $f_{\sig}^b$ and $f_{\rnd}^b$ are the fractions of signal and random-pion candidates in the Dalitz bin $b$, as determined from fits to the $(\M,\Q)$ distributions (see \cref{sec:mq-fit}). These fractions are evaluated independently for each data subset. The coefficients $\mathcal{C}_b$ account for the multinomial splitting of the total sample into the 16 Dalitz-plot bins. They are related to the ratios between the total yields in bin $-b$ and bin $+b$ by
\begin{equation}
C_b = \frac{N_{-b}}{N_{b}} = \frac{\mathcal{N}_{-b}}{\mathcal{N}_{b}}\ \frac{f_{\sig}^b}{f_\sig^{-b}}\,,
\end{equation}
where the signal yields $\mathcal{N}_{\pm b}$ are expressed in terms of the hadronic parameters $(r_b, c_b,s_b)$ and of the mixing parameters $(x,y)$ by integrating the signal PDF over the decay time (as in the denominator of \cref{eqn:pdf_sig_t_sigma_t}).

The PDF of the random-pion component, $\pdf_{\rnd}(t,\sigma_t|b)$, is obtained assuming that pairing an unrelated soft pion to a correctly reconstructed \Dz decay can only result in a fraction $f_{\rm mistag}$ of candidates being mistagged. We then write
\begin{equation}\label{eq:pdf-rnd}
\pdf_{\rnd}(t,\sigma_t|b) \propto (1-f_{\rm mistag})\mathcal{N}_b \pdf_{\sig}(t,\sigma_t|b) + f_{\rm mistag}\mathcal{N}_{-b}\pdf_{\sig}(t,\sigma_t|-b)\,.
\end{equation}
The mistag fraction is determined from a fit to data candidates populating the \Q sideband $[15,20]\mev$ to be $(42.11 \pm 0.19)\%$. This fraction is fixed in the fit to the signal region.

Simulation shows that the distribution of the remaining background varies with the Dalitz-plot bin and with the data subset. Therefore, for each bin $b$ and data subset, the PDF of the remaining background, $ \pdf_{\oth}(t,\sigma_t|b) $, is defined to be
\begin{equation}\label{eqn:pdf-oth}
\pdf_{\oth}(t,\sigma_t|b) = \pdf_{\oth}(t|\sigma_t,b)\pdf_{\oth}(\sigma_t|b) \,.
\end{equation}
The first term is the sum of a Dirac $\delta$ function and two exponential-decay components,
all convolved with the resolution function of \cref{eqn:resolution} in which the mean parameters are shifted by the bin-dependent offset $\mu_{\oth}^b$:
\begin{align}\label{eqn:pdft-oth}
\pdf_{\oth}(t|\sigma_t,b) & = (1-f_\tau^b)R(t|\sigma_t,\mu_{\oth}^b) + f_\tau^b\left[f_{\tau_1}^b\pdf_{\tau_1}(t|\sigma_t, \mu_{\oth}^b, \tau_1^b)  
 + (1-f_{\tau_1}^b)\pdf_{\tau_2}(t|\sigma_t,\mu_{\oth}^b, \tau_2^b)\right] \,,
\end{align}
with
\begin{equation}
\pdf_{\tau_i}(t|\sigma_t, \mu_{\oth}^b, \tau_i^b) \propto e^{-t/\tau_i^b}\otimes R(t|\sigma_t, \mu_{\oth}^b)\,.
\end{equation}
The PDF of $\sigma_t$ is a histogram template derived, independently for each bin $b$ and data subset, from the candidates populating the \M sideband $[1.97,2.00]\gevcc$. Parameters in $ \pdf_{\oth}(t,\sigma_t|b) $ are determined by fitting to the candidates populating the \M sideband and fixed in the fit to the signal region.

In addition, in the fit to the signal region, we fix the \Dz lifetime to the known value~\cite{pdg} and Gaussian constrain the $c_b$ and $s_b$ coefficients of the $X_b$ hadronic parameters to values obtained from the combination of the BESIII and CLEO measurements~\cite{strongphase:BESIII:2020hlg,strongphase:BESIII:2020khq}. The remaining floating parameters of the fit are $ x $, $ y $, $ r_b $, and 13 parameters of the decay-time resolution functions.

\begin{figure}[ht]
\centering
\includegraphics[width=\textwidth]{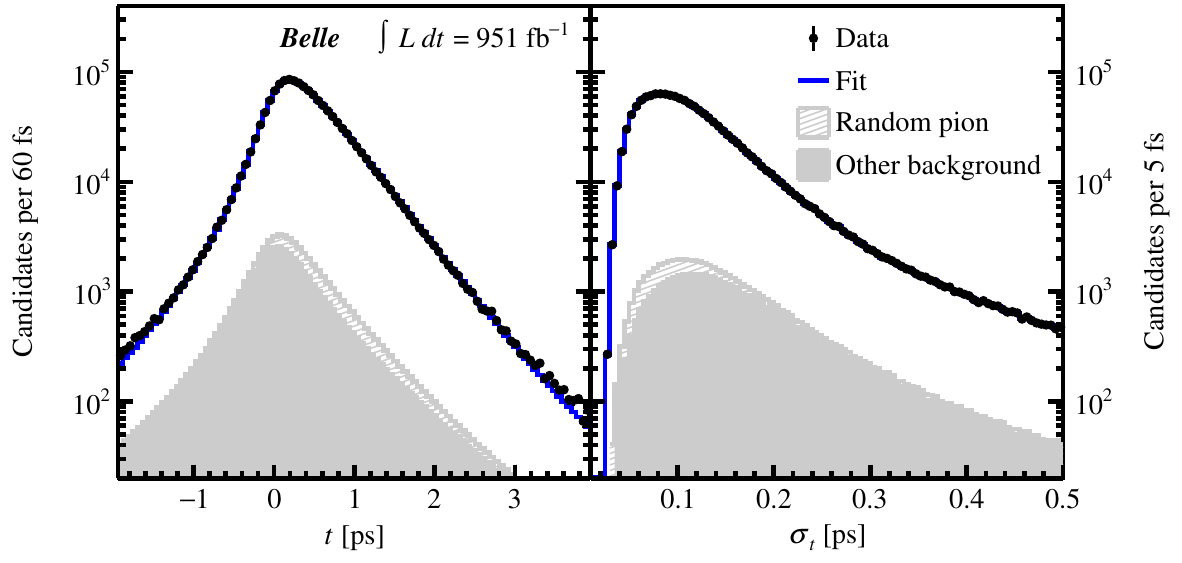}\\
\includegraphics[width=\textwidth]{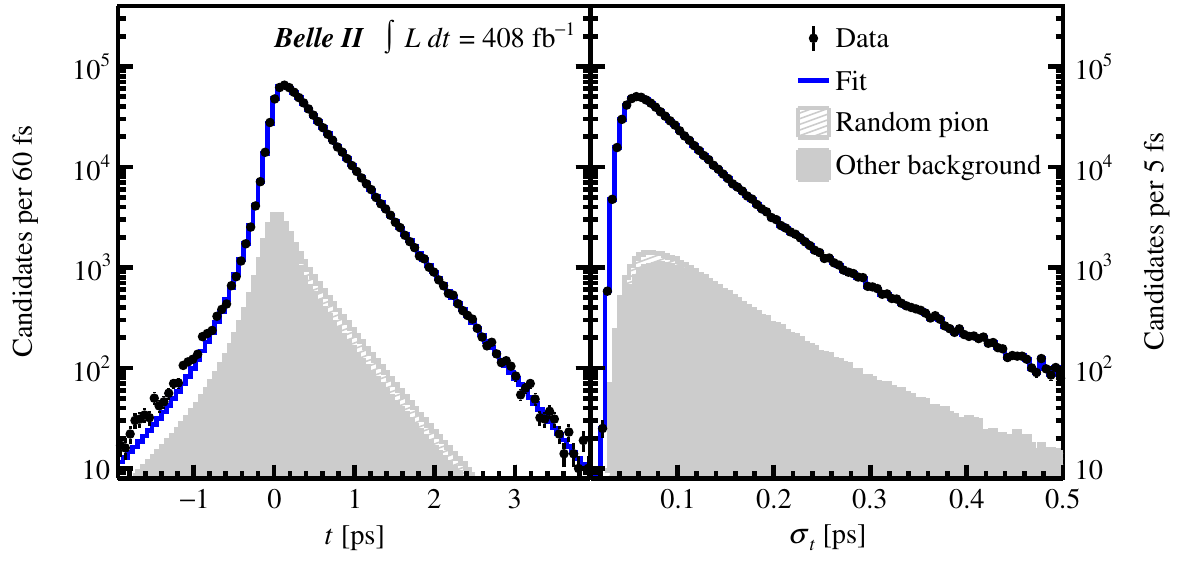}\\
\caption{Distributions, integrated over all Dalitz-plot bins, of (left) decay time and (right) decay-time uncertainty for candidates populating the signal region in (top) Belle and (bottom) Belle II data, with fit projections overlaid.\label{fig:time-dependent-fit-data-full-dalitz-plot} }
\end{figure}

The results of the fit to the data are integrated over the Dalitz-plot bins and projected over the distributions of decay time and decay-time uncertainties in \cref{fig:time-dependent-fit-data-full-dalitz-plot}. The mixing parameters are measured to be $x = (\xFit\pm\xStat) \times 10^{-3}$ and $y = (\yFit\pm\yStat) \times 10^{-3}$, where the uncertainties include the statistical uncertainties due to the limited sample size, and the contribution from the uncertainties in the external $(c_b,s_b)$ inputs. The latter are evaluated to be $0.3\times 10^{-3}$ for $x$ and $0.3\times10^{-3}$ for $y$ by computing the difference in quadrature between the uncertainties from the nominal fit and those from a fit in which the $(c_b,s_b)$ parameters are fixed to their best-fit values. The correlation between $x$ and $y$ is negligible.

\begin{figure}[ht]
\centering
\includegraphics[width=0.9\textwidth]{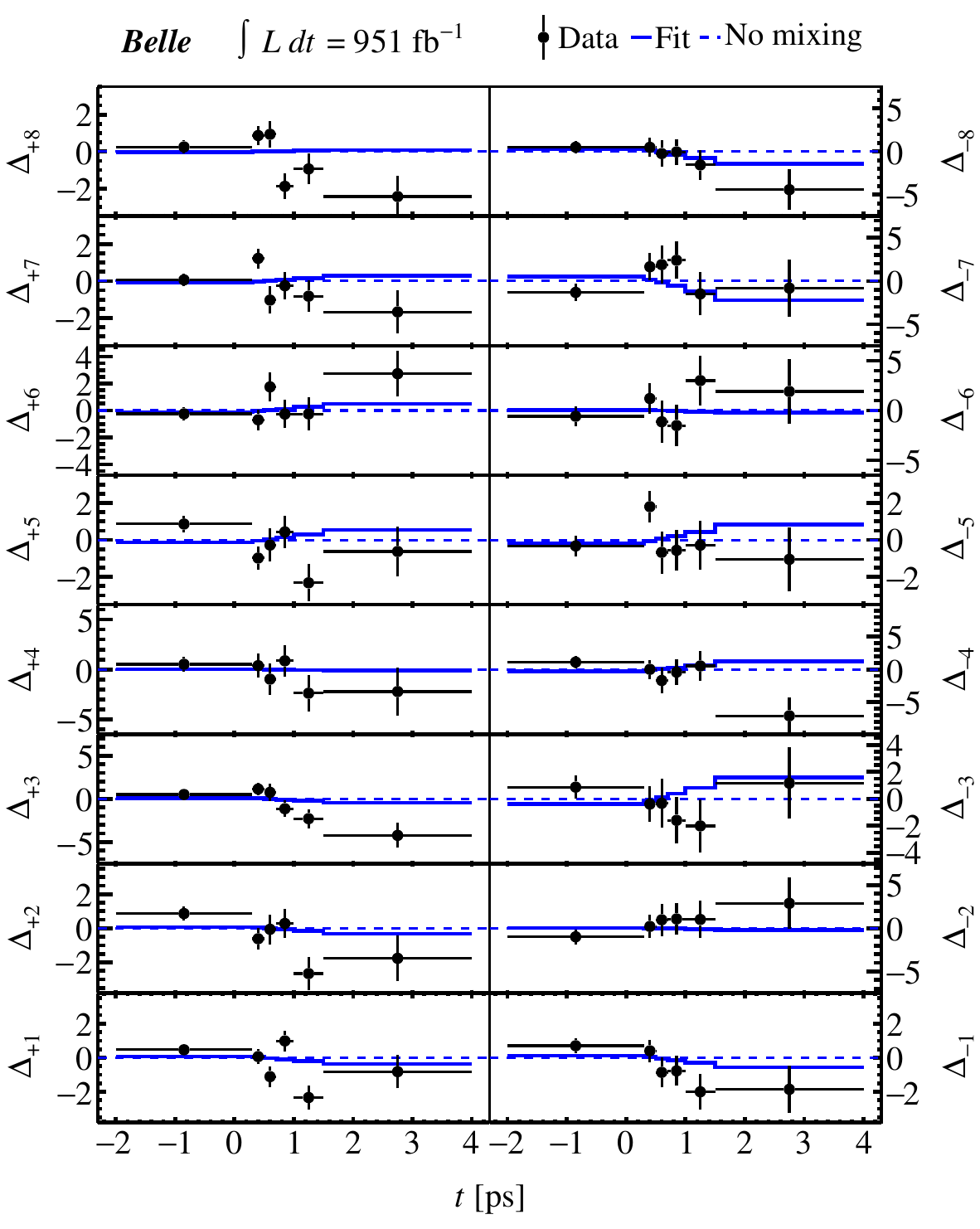}\\
\caption{Decay-time distributions of the percent relative difference between Belle data and no-mixing model, $\Delta_b=100\times(\text{data}-\text{no-mixing fit})/\text{no-mixing fit}$, in each Dalitz-plot bin $b$, with fit projections overlaid.\label{fig:mixing-visualization-belle}}
\end{figure}

\begin{figure}[ht]
\centering
\includegraphics[width=0.9\textwidth]{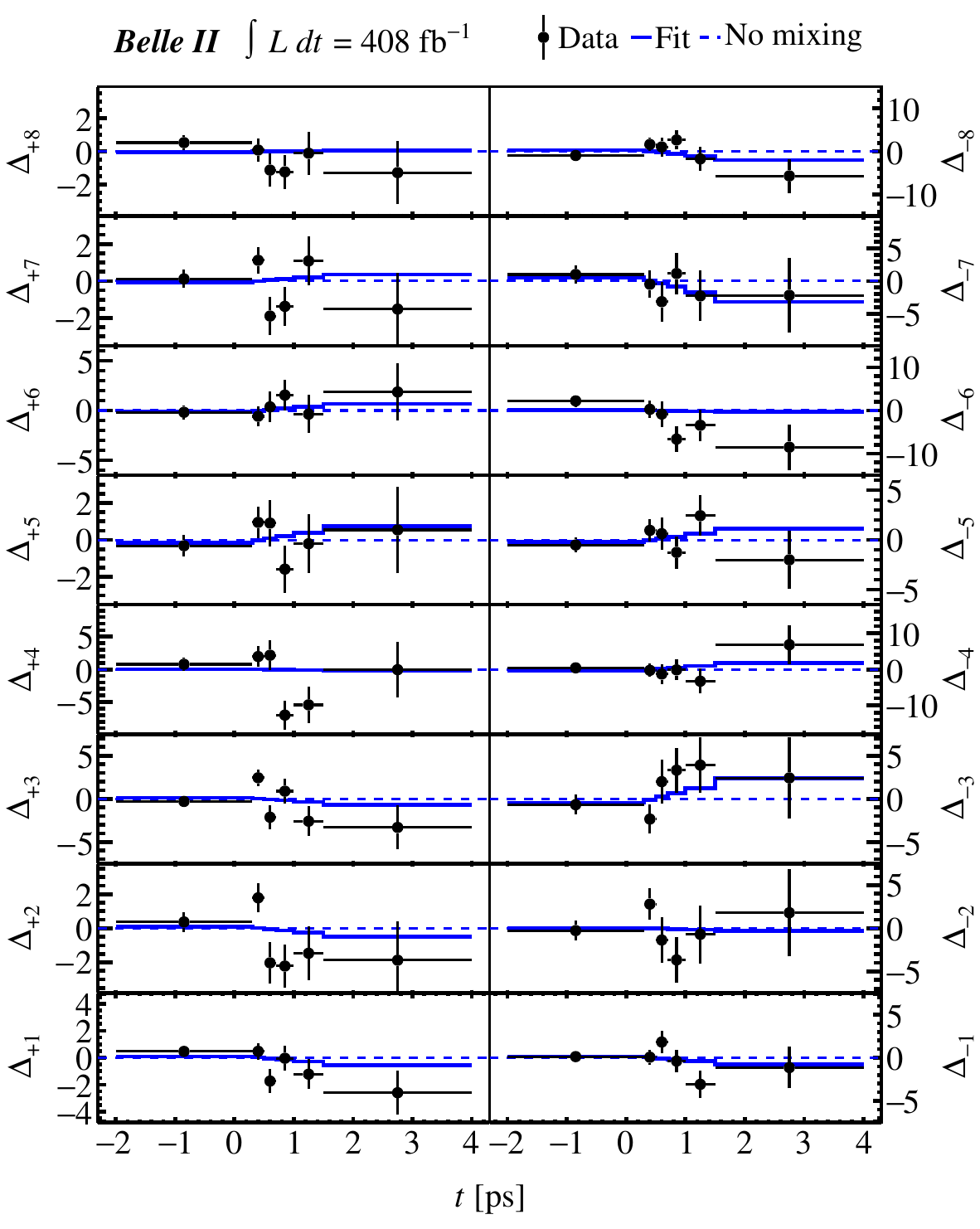}\\
\caption{Decay-time distributions of the percent relative difference between Belle II data and no-mixing model, $\Delta_b=100\times(\text{data}-\text{no-mixing fit})/\text{no-mixing fit}$, in each Dalitz-plot bin $b$, with fit projections overlaid.\label{fig:mixing-visualization-belle2}}
\end{figure}

A fit to the data assuming $x=y=0$ is performed to evaluate the consistency of the data with the no-mixing hypothesis. \Cref{fig:mixing-visualization-belle,fig:mixing-visualization-belle2} show the decay-time distributions of the relative difference between data and no-mixing model in each Dalitz-plot bin $b$ for Belle and Belle II, respectively, with mixing-allowed fit projection overlaid. From the ratio between the likelihood values of the mixing-allowed and no-mixing fits we find the data to be consistent with the no-mixing hypothesis with a $p$ value of 0.7\% (corresponding to approximately 2.7 Gaussian standard deviations).

\section{Systematic uncertainties and cross-checks}\label{sec:systematics}
We consider the following sources of systematic uncertainties: nonuniform efficiency across the Dalitz plot; resolution on the Dalitz-plot coordinates; modeling of the decay-time resolution; modeling of the background; uncertainty in the measured mistag rate; and uncertainty in the input \Dz lifetime. \Cref{tab:systematics_absolute} lists the estimated uncertainties. The total uncertainty is the sum in quadrature of the individual components.

\begin{table}[t]
\centering
\caption{Systematic uncertainties (in units of $10^{-3}$).\label{tab:systematics_absolute}}
\begin{tabular}{lrr}
\hline
Source & \multicolumn{2}{c}{\quad Uncertainties}\\
       & \multicolumn{1}{c}{$\quad x$} & \multicolumn{1}{c}{$\quad y$} \\
\hline
Nonuniform Dalitz-plot efficiency  & $0.10$ & $0.03$ \\
Dalitz-plot resolution  & $0.04$ & $0.12$ \\
Decay-time resolution model  & $0.06$ & $0.04$ \\
Background model        & $0.38$ & $0.30$ \\
Mistag rate             & $0.12$ & $0.04$ \\
Input \Dz lifetime    & $0.01$ & $0.02$ \\
\hline
Total & $\xSyst\phantom{0}$ & $\ySyst\phantom{0}$ \\
\hline
\end{tabular}
\end{table}

The analysis procedure is validated with pseudo-experiments generated using the $\Dz\to\KS\pip\pim$ amplitude model from Ref.~\cite{delAmoSanchez:2010xz} and with the assumed resolution and background models. No bias is observed in the estimated mixing parameters or in their uncertainties.

Pseudo-experiments are also used to evaluate the systematic uncertainties resulting from the nonuniform Dalitz-plot efficiency, the resolution on the Dalitz-plot variables, the decay-time resolution model, and the background models. For each case, we generate pseudo-experiments that mimic alternative realistic models (derived from simulation) and fit them using the nominal models. The resulting average deviations of the measured mixing parameters from the generated values is assigned as the systematic uncertainty. The largest systematic uncertainty arises from the background model. The model neglects $\lesssim0.1\%$ contributions from partially or misreconstructed \Dz decays (e.g., $\Dz\to\KS\pim\mu^+\nu_\mu$, $\Dz\to\KS K^\pm\pi^\mp$, $\Dz\to2\pip2\pim$, $\Dz\to\KS\KS$) populating the signal region. When simulated in our pseudo-experiments, the neglected backgrounds bias the background fractions estimated by the \M versus \Q fit and affect the reliability of $t$ versus $\sigma_t$ model obtained from sideband data. In turn, these effects result in biases on the mixing parameters.

While fitting to the data the mistag rate is fixed to the value, $(42.11 \pm 0.19)\%$, measured from the fit to the \Q sideband. We refit the data with the mistag fraction varied by $\pm0.19\%$ and take the largest observed differences with respect to the nominal results as the systematic uncertainty.

The relative uncertainty on the input value of the \Dz lifetime, $2.4\times10^{-3}$~\cite{pdg}, propagates into a relative uncertainty on the mixing parameters and makes a small contribution to the total systematic uncertainties. Other effects, such as possible biases from multiple \Dstarp candidates in an event, are also investigated and found to be negligible.

\begin{figure}[ht]
\centering
\includegraphics[width=0.5\textwidth]{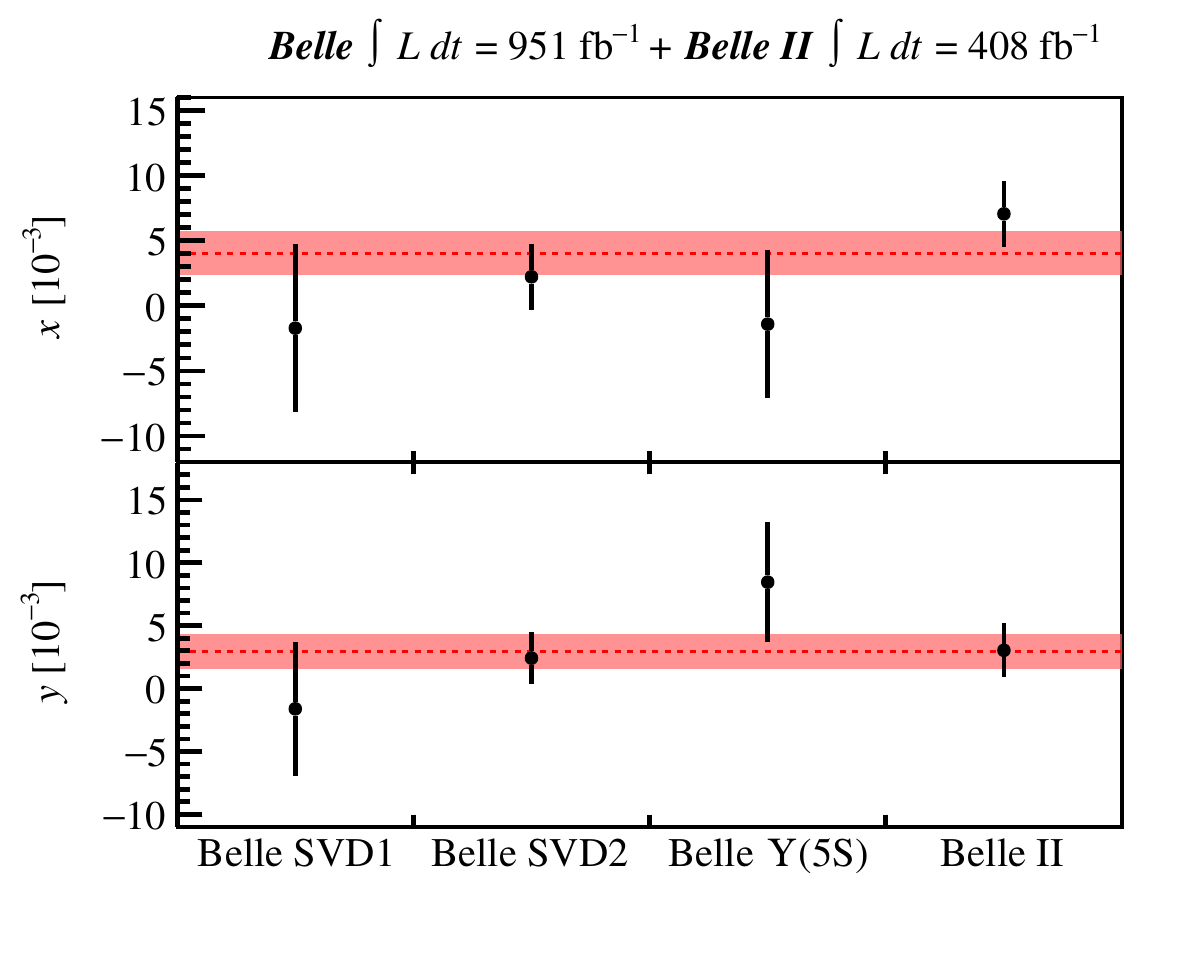}\\
\caption{Mixing parameters from independent fits to subsets of the data. The dashed lines and the bands represent the nominal result from the simultaneous fit of all the subsets. Uncertainties are statistical only.\label{fig:subsets}}
\end{figure}

Finally, the stability of the results is checked by performing the measurement in independent subsets of the data defined according to data-taking conditions, \KS flight length, \Dz momentum, and \Dz polar angle. In all cases, we obtain results in agreement with each other, and with the results from the full sample. \Cref{fig:subsets} shows the consistency of the mixing parameters measured in different data subsets.

\section{Conclusions}
In summary, we perform a model-independent measurement of the \Dz-\Dzb mixing parameters using $2.05\times10^6$ \Dstarp-tagged $\Dz\to\KS\pip\pim$ decays reconstructed in Belle and Belle II data samples, which have integrated luminosities of 951\invfb and 408\invfb, respectively. Assuming \CP symmetry, the mixing parameters are measured to be
\begin{align}
x & = (\xFit\pm\xStat\pm\xSyst) \times 10^{-3}\,, \\
y & = (\yFit\pm\yStat\pm\ySyst) \times 10^{-3}\,,
\end{align}
where the first uncertainties are statistical and the second are systematic. The statistical uncertainties also include the contribution from the uncertainties in the external strong-phase inputs~\cite{strongphase:BESIII:2020hlg,strongphase:BESIII:2020khq}. These results are consistent with previous determinations. They are approximately 20\% and 14\% more precise, and have significantly smaller systematic uncertainties, than the model-dependent Belle measurement presented in Ref.~\cite{Peng:2014oda}. 
 
\begin{acknowledgements}
%
%
This work, based on data collected using the Belle II detector, which was built and commissioned prior to March 2019,
and data collected using the Belle detector, which was operated until June 2010,
was supported by
Higher Education and Science Committee of the Republic of Armenia Grant No.~23LCG-1C011;
Australian Research Council and Research Grants
No.~DP200101792, %
No.~DP210101900, %
No.~DP210102831, %
No.~DE220100462, %
No.~LE210100098, %
and
No.~LE230100085; %
Austrian Federal Ministry of Education, Science and Research,
Austrian Science Fund
No.~P~34529,
No.~J~4731,
No.~J~4625,
and
No.~M~3153,
and
Horizon 2020 ERC Starting Grant No.~947006 ``InterLeptons'';
Natural Sciences and Engineering Research Council of Canada, Compute Canada and CANARIE;
National Key R\&D Program of China under Contract No.~2022YFA1601903,
National Natural Science Foundation of China and Research Grants
No.~11575017,
No.~11761141009,
No.~11705209,
No.~11975076,
No.~12135005,
No.~12150004,
No.~12161141008,
No.~12475093,
and
No.~12175041,
and Shandong Provincial Natural Science Foundation Project~ZR2022JQ02;
the Czech Science Foundation Grant No.~22-18469S 
and
Charles University Grant Agency project No.~246122;
European Research Council, Seventh Framework PIEF-GA-2013-622527,
Horizon 2020 ERC-Advanced Grants No.~267104 and No.~884719,
Horizon 2020 ERC-Consolidator Grant No.~819127,
Horizon 2020 Marie Sklodowska-Curie Grant Agreement No.~700525 ``NIOBE''
and
No.~101026516,
and
Horizon 2020 Marie Sklodowska-Curie RISE project JENNIFER2 Grant Agreement No.~822070 (European grants);
L'Institut National de Physique Nucl\'{e}aire et de Physique des Particules (IN2P3) du CNRS
and
L'Agence Nationale de la Recherche (ANR) under grant ANR-21-CE31-0009 (France);
BMBF, DFG, HGF, MPG, and AvH Foundation (Germany);
Department of Atomic Energy under Project Identification No.~RTI 4002,
Department of Science and Technology,
and
UPES SEED funding programs
No.~UPES/R\&D-SEED-INFRA/17052023/01 and
No.~UPES/R\&D-SOE/20062022/06 (India);
Israel Science Foundation Grant No.~2476/17,
U.S.-Israel Binational Science Foundation Grant No.~2016113, and
Israel Ministry of Science Grant No.~3-16543;
Istituto Nazionale di Fisica Nucleare and the Research Grants BELLE2;
Japan Society for the Promotion of Science, Grant-in-Aid for Scientific Research Grants
No.~16H03968,
No.~16H03993,
No.~16H06492,
No.~16K05323,
No.~17H01133,
No.~17H05405,
No.~18K03621,
No.~18H03710,
No.~18H05226,
No.~19H00682, %
No.~20H05850,
No.~20H05858,
No.~22H00144,
No.~22K14056,
No.~22K21347,
No.~23H05433,
No.~26220706,
and
No.~26400255,
and
the Ministry of Education, Culture, Sports, Science, and Technology (MEXT) of Japan;  
National Research Foundation (NRF) of Korea Grants
No.~2016R1-D1A1B-02012900,
No.~2018R1-A6A1A-06024970,
No.~2021R1-A6A1A-03043957,
No.~2021R1-F1A-1060423,
No.~2021R1-F1A-1064008,
No.~2022R1-A2C-1003993,
No.~2022R1-A2C-1092335,
No.~RS-2023-00208693,
No.~RS-2024-00354342
and
No.~RS-2022-00197659,
Radiation Science Research Institute,
Foreign Large-Size Research Facility Application Supporting project,
the Global Science Experimental Data Hub Center, the Korea Institute of
Science and Technology Information (K24L2M1C4)
and
KREONET/GLORIAD;
Universiti Malaya RU grant, Akademi Sains Malaysia, and Ministry of Education Malaysia;
Frontiers of Science Program Contracts
No.~FOINS-296,
No.~CB-221329,
No.~CB-236394,
No.~CB-254409,
and
No.~CB-180023, and SEP-CINVESTAV Research Grant No.~237 (Mexico);
the Polish Ministry of Science and Higher Education and the National Science Center;
the Ministry of Science and Higher Education of the Russian Federation
and
the HSE University Basic Research Program, Moscow;
University of Tabuk Research Grants
No.~S-0256-1438 and No.~S-0280-1439 (Saudi Arabia);
Slovenian Research Agency and Research Grants
No.~J1-9124
and
No.~P1-0135;
Ikerbasque, Basque Foundation for Science,
the State Agency for Research of the Spanish Ministry of Science and Innovation through Grant No. PID2022-136510NB-C33,
Agencia Estatal de Investigacion, Spain
Grant No.~RYC2020-029875-I
and
Generalitat Valenciana, Spain
Grant No.~CIDEGENT/2018/020;
the Swiss National Science Foundation;
The Knut and Alice Wallenberg Foundation (Sweden), Contracts No.~2021.0174 and No.~2021.0299;
National Science and Technology Council,
and
Ministry of Education (Taiwan);
Thailand Center of Excellence in Physics;
TUBITAK ULAKBIM (Turkey);
National Research Foundation of Ukraine, Project No.~2020.02/0257,
and
Ministry of Education and Science of Ukraine;
the U.S. National Science Foundation and Research Grants
No.~PHY-1913789 %
and
No.~PHY-2111604, %
and the U.S. Department of Energy and Research Awards
No.~DE-AC06-76RLO1830, %
No.~DE-SC0007983, %
No.~DE-SC0009824, %
No.~DE-SC0009973, %
No.~DE-SC0010007, %
No.~DE-SC0010073, %
No.~DE-SC0010118, %
No.~DE-SC0010504, %
No.~DE-SC0011784, %
No.~DE-SC0012704, %
No.~DE-SC0019230, %
No.~DE-SC0021274, %
No.~DE-SC0021616, %
No.~DE-SC0022350, %
No.~DE-SC0023470; %
and
the Vietnam Academy of Science and Technology (VAST) under Grants
No.~NVCC.05.12/22-23
and
No.~DL0000.02/24-25.

These acknowledgements are not to be interpreted as an endorsement of any statement made
by any of our institutes, funding agencies, governments, or their representatives.

We thank the SuperKEKB team for delivering high-luminosity collisions;
the KEK cryogenics group for the efficient operation of the detector solenoid magnet and IBBelle on site;
the KEK Computer Research Center for on-site computing support; the NII for SINET6 network support;
and the raw-data centers hosted by BNL, DESY, GridKa, IN2P3, INFN, 
PNNL/EMSL, 
and the University of Victoria.
 \end{acknowledgements}

\bibliographystyle{belle2}
\providecommand{\href}[2]{#2}\begingroup\raggedright\endgroup

\end{document}